\documentclass[preprint,prb,showpacs,preprintnumbers,amsmath,amssymb]{revtex4}
\usepackage{graphicx}
\usepackage{dcolumn}
\usepackage{color}
\usepackage{bm}
\newcommand{\ve}{\varepsilon}

\begin{document}

\title{Current-induced interactions of multiple domain walls in magnetic quantum wires}

\author{N. Sedlmayr$^1$,}
\thanks{nicholas.sedlmayr@physik.uni-halle.de}
\author{V. K. Dugaev$^2$, and J. Berakdar$^1$}
\affiliation{$^1$
Department of Physics, Martin-Luther-Universit\"at Halle-Wittenberg,
Heinrich-Damerow-Str. 4, 06120, Halle, Germany\\
$^2$ Department of Physics, Rzesz\'ow University of Technology,
al. Powsta\'nc\'ow Warszawy 6, 35-959 Rzesz\'ow, Poland}
\affiliation{Department of Physics and CFIF, Instituto Superior T\'ecnico,
TU Lisbon, av. Rovisco Pais, 1049-001, Lisbon, Portugal}

\date{\today}

\begin{abstract}
We show that an applied charge current in a magnetic  nanowire containing domain walls (DWs)
results in an interaction between DWs mediated by spin-dependent
interferences of the scattered carriers.
The energy and torque associated with this interaction show an oscillatory behaviour
as a function of the  mutual DWs orientations and separations, thus affecting the DWs'
arrangements and shapes.
Based on the derived  DWs interaction energy and torque we calculate DW dynamics and
uncover potential applications of interacting DWs as a tunable nano-mechanical oscillator. We also discuss the effect of impurities on the DW
interaction.
\end{abstract}


\maketitle

\section{Introduction}

Domain walls (DWs), i.e.~regions of noncollinearity separating
areas of different homogenous magnetization directions, are
important from a fundamental and application point of view \cite{yamanouchi04,yamaguchi04,parkin08}.
This is particularly the case at low dimensions,  as in
 magnetic nanowires   carriers turn out to couple strongly with
DWs \cite{klaui08}
leading to a marked influence on the wire's transport properties, e.g.  DW magnetoresistances in
the range of 1000\% were reported \cite{ebels00,chopra02,ruster03} .
As this coupling is associated with a change of the carriers' spin, it results in
a current-induced spin torque acting on the DW and consequently in a
current-induced DW motion \cite{yamanouchi04,yamaguchi04}.
Based on these facts magnetic nanowires with a series of DWs can be utilized as
a ``racetrack DW memory'' \cite{parkin08}. The DWs'  motion is current-controlled;
 DWs separated by rather small distances are addressable thus allowing for a high memory density.

In another context it is established that strong carrier scattering and interference
results in  long-range interactions
between impurities on metal surfaces.
This interaction governs the impurities geometric arrangements and growth
\cite{surface1,surface2,surface3,surface4,surface5,surface6}.
The question of whether and how the carriers' spin dependent scattering
  mediates interactions between DWs is still outstanding and should be addressed
here. Clearly, the answer is of vital importance for a high-density nanowire-based racetrack memory
and adds a new twist on interference-mediated interactions.
 We focus on the  { current-induced} part of the coupling between neighboring DWs
 in a magnetic nanowire.
 Based on our results
  we identify the following  mechanism of the DWs coupling: Upon scattering from the first DW
 a carrier spiral spin density builds up. This acts as a spatially non-uniform  torque on
 the second DW whose energetically stable shape and position
  show  therefore a non-uniform dependence on the distance from the first DW. This
is different from the spin-torque transfer
in bulk spin valve systems \cite{slonczewski96,berger96,barnas05}
or magnetic tunnel junctions \cite{slonczewski89,theodonis06} insofar as in our case the  DWs spatial arrangement, in addition to the magnetization direction, is current controlled \footnote{Also note, in reduced  dimensions, e.g. a constrained nanowires,  the propagation of the
transverse component of the torque is strongly enhanced; in contrast this component is
  suppressed in bulk systems \cite{stiles02}.}.
We develop  a theoretical framework to calculate  the  DWs current-induced
effective potential  and find it oscillates with the  distance of DWs  and their
 mutual polarization directions. This interaction we employ to study the  DWs  dynamics.
As an application
 we propose the use of this new effect as a tunable, current-driven  two-DW magnetic nano-oscillator \cite{katine00,kiselev03,kaka05};
 with a radiation emission dependent on the DWs positions in the various possible stable configurations.

\section{Theoretical model}

We consider a magnetic nanowire with two DWs when an electric current $I$
is transmitted through the wire (a schematic is shown in Fig.~\ref{deltae2d}).
When the  distance  $z_0$ between DWs is larger than
the phase coherence length $L_\phi$, DWs act as independent scatterers. For
$z_0\lesssim L_\phi$  the current transmission mediates
DW coupling. For definiteness,
we assume that one of the DWs (located at $z=0$) is pinned, e.g. by a geometric constriction,
 and concentrate on the effect
of the current on the second DW, initially (i.e., for $I=0$) located  at $z=z_0$. For
$I=0$ each DW has  an extension $L$.
The transverse dimensions of the wire should be smaller than the exchange length and the
Fermi wave length of the carriers, a situation realizable for magnetic semiconductors.
The Hamiltonian $\bar H$ of independent carriers  coupled (with a coupling constant $J$)
to a spatially non-uniform
magnetization (DWs) profile ${\bf M}(z)$ is modeled by  (we use units with $\hbar=1$)
\begin{eqnarray}
\bar H=\int dz\, a_\alpha ^\dag (z)
\left[ -\frac{\partial _z^2}{2m}\, \delta _{\alpha\beta}
-J\, \bm{\sigma}_{\alpha\beta}\cdot {\bf M}(z)\right] a_\beta (z),
\end{eqnarray}
where $a^\dag_{\alpha}$ and $a_\alpha $ are the creation and annihilation operators
of electrons with spin $\alpha $.
Applying a local gauge transformation $T(z)$
\cite{kor,tatara,dugaev1}
we obtain instead of the nonuniform magnetization
 a Zeeman splitting term and a spin-dependent
spatially varying potential
$U_{\alpha\beta}(z)$, which for $k_F L\gtrsim 1$
can be treated perturbatively  \cite{kor,tatara,dugaev1,ijmp}
($k_F$ is the Fermi wave vector). Note, for a sharp domain wall, i.e. for $k_F L< 1$, the formalism of Ref.\cite{dugaev3} can be adopted.
  $T(z)$ is obtained from the requirement
$T^\dag(z)\, {\bm \sigma} \cdot {\bf n}(z)\, T(z)
=\sigma ^z$, where ${\bf n}$ is the unit vector along ${\bf M}$,
${\bf M}(z)=M\, {\bf n}(z)$.
The transformed Hamiltonian  $H=T^\dag(z)\, {\bar H} \, T(z)
$ reads
\begin{eqnarray}
\label{hamiltonian}
H=\int dz\, a_\alpha^\dag(z)\left[
-\frac{\partial _z ^2}{2m}\, \delta_{\alpha\beta}
+U_{\alpha\beta}(z)
-JM\sigma^z_{\alpha\beta}\right ] a_\beta(z) \hskip0.2cm
\end{eqnarray}
with the perturbation given by
\begin{eqnarray}
U(z)=-\frac{1}{2m}\left[ 2A\, \partial _z+
(\partial _z A)+A^2\right] ,
\end{eqnarray}
and
$A(z)=T^\dag (z)\, \nabla _z\, T(z)$ is a gauge potential.
For a wire with two DWs we  parametrize
the magnetization profile by the angles $\varphi (z)$ and $\theta (z)$ (cf. Fig.\ref{deltae2d})
\begin{eqnarray}
{\bf n}(z)&=&\big( \cos \theta \, \sin \varphi ,\;
\sin \theta \, \sin \varphi ,\;
\cos \varphi \big) ,\label{magnetization1}\\
\varphi(z)
&=&\underbrace{\cos^{-1}\big(\tanh\big[\frac{z}{L}\big]\big)}_{=-\varphi_1(z)} +\underbrace{\cos^{-1}\big(\tanh\big[\frac{z-z_0}{L}\big]\big)}_{=-\varphi_2(z)}.
\label{magnetization2}
\end{eqnarray}
(See reference \cite{thiaville06} and references therein.) The angle $\theta(z)$ describes the relative orientation between the wall pinned at $z=0$ and the
other situated around $z=z_0$.
We set  $\theta_1 $ to zero at the first wall  and $\theta_2 =\theta_0$ around the second
 (see Fig.~\ref{deltae2d} ).
For $\theta_0=\pi $  the walls are antialigned.
For $z_0\leq L$ DWs may merge, hence we consider the case $z_0> L$ for which
 we may write $U(z)\approx U_1(z)
+U_2(z)$, where ($j=1,2$)
\begin{eqnarray}
U_j(z)&=&
\frac{[\varphi_j'(z)]^2}{8m}+i\sigma ^y
\bigg[\frac{\varphi_j''(z)}{4m}+\frac{\varphi_j'(z)\, \partial_z}{2m}\bigg]
\cos \theta_j
\nonumber\\&&
-i\sigma ^x \bigg[\frac{\varphi_j''(z)}{4m}+\frac{\varphi_j'(z)\, \partial_z}{2m}\bigg]
\sin \theta_j.
\label{u2}
\end{eqnarray}
This approach is generalizable to any number of DWs,
which are sufficiently far apart.
As shown in Refs.\cite{tatara,dugaev1,ijmp} for a single DW, for $k_FL\geq 1$,
i.e. when $\mathbf M (z)$ hardly varies within $k_F^{-1}$(adiabatic DW),
the terms in Eq.~(\ref{u2}) proportional to $\varphi_1''(z)$  are negligibly small
  and a perturbative approach is appropriate for treating
  the  electron scattering  from the DWs potential (Eq.~(\ref{u2}))
  \footnote{In fact, as shown in \cite{ijmp}, this approach is justifiable even for
   $k_FL=1$ \cite{ijmp}.}.
Assuming $\psi ^0(z)$ to be the wave function of an independent electron with energy $\varepsilon $
in the wire without the DWs, we
find the first-order correction
due to the perturbation $U_1(z)$, i.e. due to scattering from the first DWs, as
\begin{eqnarray}
\delta \psi _\ve (z)=\int_{-\infty}^\infty dz'\,
G_\ve(z,z')\, U_1(z')\, \psi ^0(z').
\end{eqnarray}
The Green's function $G_\varepsilon $ corresponds to the unperturbed
Hamiltonian with $U(z)=0$. It is diagonal in spin space with elements
\begin{eqnarray}\label{freeg}
G_{\ve \sigma }(z,z')=-\frac{im}{k_\sigma }\, e^{ik_\sigma |z-z'|},
\end{eqnarray}
where $k_\sigma \approx k^0_\sigma +\frac{i}{2\tau_\sigma}\frac{m}{k^0_\sigma}$
for lifetimes $\tau_\sigma \gg \ve_F^{-1}$, and
$k^0_\sigma =\left[ 2m\, (\ve+\mu\pm JM)\right] ^{1/2}$. Hence
\begin{eqnarray}
\delta \psi _{\ve \uparrow }(z)
=\int_{-\infty}^\infty dz'
\begin{pmatrix}
-\frac{i}{8k_\uparrow }e^{ik_\uparrow |z-z'|}[\varphi'_1(z)]^2\, e^{ik_\uparrow z'}
\\
-\frac{k_\uparrow }{2k_\downarrow }e^{ik_\downarrow |z-z'|}\, \varphi'_1(z)\, e^{ik_\uparrow z'}
\end{pmatrix}
\end{eqnarray}
and
\begin{eqnarray}
\delta \psi _{\ve \downarrow }(z)
=\int_{-\infty}^\infty dz'
\begin{pmatrix}
\frac{k_\downarrow }{2k_\uparrow }e^{ik_\uparrow |z-z'|}\, \varphi'_1(z)]\, e^{ik_\downarrow z'}
\\
-\frac{i}{8k_\downarrow }e^{ik_\downarrow |z-z'|}\, [\varphi'_1(z)]^2\, e^{ik_\downarrow z'}
\end{pmatrix}
\end{eqnarray}
for incoming electrons of spin up and down, respectively.

The interaction energy of the two DWs due to the single scattered state
$\psi _{\ve \sigma }(z)=\psi ^0_{\ve \sigma }(z)+\delta \psi _{\ve \sigma }(z)$ is
calculated as
\begin{eqnarray}
\Delta E_\sigma =\int_{-\infty}^\infty dz\,
\delta\psi^\dagger_{\ve\sigma}(z)\,
U_2(z)\, \delta\psi_{\ve\sigma}(z).
\end{eqnarray}
Summing up the contributions of all scattering states in the energy range
between $\ve_F$ and $\ve_F+e\Delta\phi /2$, for an applied voltage
$e\Delta\phi/2\ll \ve_F$, we obtain  the current-induced coupling of the DWs as
\begin{eqnarray}
\Delta E=\frac{e\Delta\phi}{\sqrt{2}\pi}
\bigg(\frac{\Delta E_\uparrow }{v_\uparrow }
+\frac{\Delta E_\downarrow}{v_\downarrow }\bigg),
\label{deltaeeq}
\end{eqnarray}
where $v_\sigma =k^0_\sigma /m$ is the velocity of electrons at the Fermi level.

\section{Numerical examples}

\begin{figure}
\includegraphics[width=0.6\textwidth,angle=270]{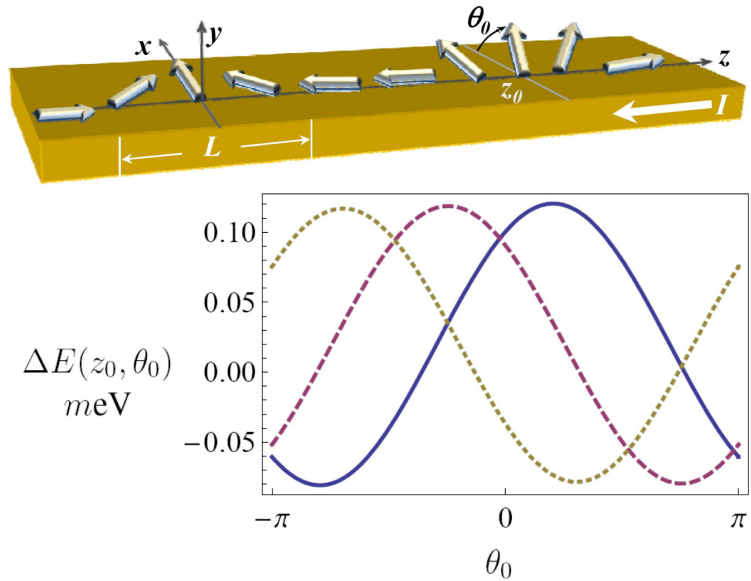}
\caption[$\Delta E$]{(Color online) Top panel:
A schematics  showing the DWs magnetization profile (thick arrows).
 $L$ is the DW width, $z_0$ and $\theta_0$ are respectively
  the DW position and orientation with respect to the DW at $z=0$, $I$ is the current direction.
 Lower panel: Interaction energy $\Delta E(z_0,\theta_0)$ as
a function of $z_0$ and $\theta_0$.
Solid  curve is for $z_0=30$~nm, the dashed  is for $z_0=37.5$~nm,
and the dotted  is for $z_0=45$~nm. }
\label{deltae2d}
\end{figure}

Magnetic semiconductors
\cite{ruster03,sugawara08} are most favorable for a sizable effect, for metallic wires
the 1D limit is also within reach \cite{claessen}. Here  we use in the
 numerical calculations similar parameters as in Ref.\onlinecite{ruster03}, i.e.
 $\lambda_F=6$~nm; a mean free path of $l=500$~nm; an effective mass of $m=0.5m_e$ ($m_e$ is free electron mass); $L=\lambda_F$; $JM=15$~meV; $\ve_F=83.7$~meV; and $e\Delta\phi=0.1\ve_F$.
 \footnote{We note that the conditions for our  theory to be applicable is that the mean free path $l$
 should be larger than the distance $L$ between the DWs.
 A carrier wave length larger than $l$ means a complete localization (Ioffe-Regel criteria of localization),
 a case which is not of interest here. On the other hand,
 in a magnetic semiconductor, $l$ can be enhanced by lowering
 the acceptors density (and/or ordering them).}
 The width of the wall may well be on the atomic size in the
 presence of constrictions\cite{bruno99,pietzsch00,ebels00}, i.e. well below the DW lengths  in bulk materials.
 In such a situation, the DW-interaction increases due to the strongly enhanced
 DW scattering \cite{jb_du,jb_du2,jb_du3,jb_du4}.
The interaction energy  Figure \ref{deltae2d}  depends periodically on the DWs
 mutual angle $\theta_0$ and  distance $z_0$,
which results in an oscillating motion of the DW along the axis $z$
as well as an oscillating direction of DW polarization.

Now we focus on the effect of DW scattering on the electron
spin density, leading to a nonequilibrium spin accumulation and to
a spin torque acting on the wall.
Subsequently, we study the dynamics of the DW related to the DW coupling.

The spin-density due to the single transmitted wave of spin
$\sigma$ is
\begin{eqnarray}
{\bf S}_\sigma (z)
=\psi ^\dag _{\ve\sigma}(z)\, T(z)\,
\bm{\sigma}\, T^\dag (z) \, \psi _{\ve\sigma}(z),
\end{eqnarray}
and the total current-induced spin density is \cite{dugaev2}
\begin{eqnarray}\label{spin}
{\bf S}(z)=\frac{e\, \phi}{2\pi}
\bigg( \frac{{ \bf S}_{\uparrow}}{v_\uparrow }
+\frac{{\bf S}_{\downarrow }}{v_\downarrow } \bigg) .
\end{eqnarray}
We find that the correction to the spin density follows
the magnetization profile with additional Friedel oscillations,
which are a superposition of two waves with periods $k^{-1}_{F\uparrow }$ and
$k^{-1}_{F\downarrow }$. The oscillations in the spin density are smaller in
magnitude than the overall spin density profile and decay with increasing $z$.

\begin{figure}
\includegraphics[width=0.45\textwidth]{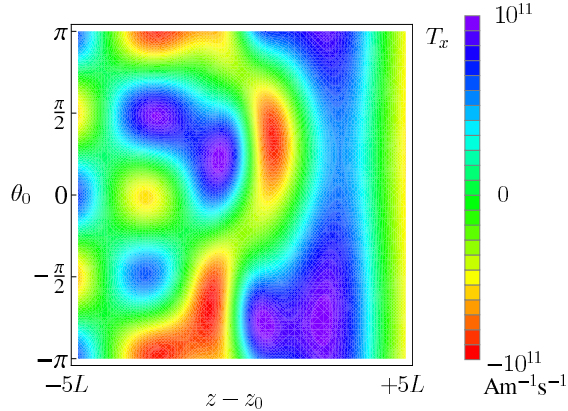}
\caption[$T_x(z,\theta_0)$: Spin Torque Density Plot Correction]
{(Color online) The $x$-component
of the current-induced spin torque, as defined in Eq.~\ref{torque},
acting at the second domain wall as a function of $z$ and $\theta_0$.
}\label{densitytorx}
\end{figure}

\begin{figure}
\includegraphics[width=0.45\textwidth]{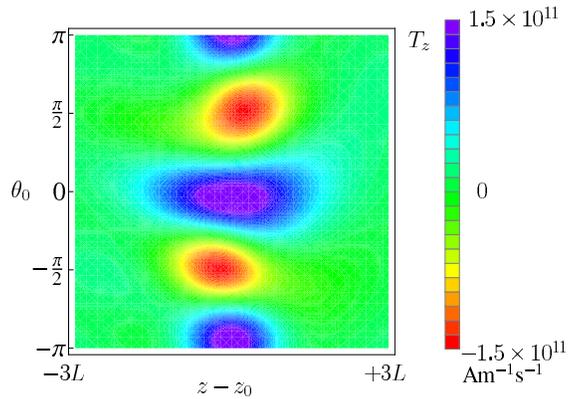}
\caption[$T_x(z,\theta_0)$: Spin Torque Density Plot Correction]
{(Color online) The $z$-component
of the current-induced spin torque
around the second domain wall as a function of $z$ and $\theta_0$.
}\label{densitytorz}
\end{figure}

We calculate the current-induced torque acting on the second DW at $z$ from
\begin{eqnarray}
\label{torque}
\Delta {\bf T}(z,z_0,\theta_0)
=-\frac{\gamma J}{\sigma_{cs}}{\bf M}(z,z_0,\theta_0)\times\Delta {\bf S}(z,z_0,\theta_0),
\end{eqnarray}
where
$\gamma=g\mu_B$, $g$ is the Land\'e factor and $\mu_B$ is the Bohr magneton.
We assumed a thin nanowire with a cross section  of $\sigma_{cs}=100\times20$~nm${}^2$ as in Ref.~\cite{ruster03}. In Eq.~(15)
$\Delta {\bf S}$ is the correction to the electron spin density due to
scattering.
The calculated torque on the second DW is shown in Figs.~\ref{densitytorx}
and \ref{densitytorz},
where $z_0=50\, L$ and $M\approx5.56\times10^{4}$Am${}^{-1}$ were used\cite{sugawara08}. The correction to the spin torque shows that the
force upon the DW depends strongly on their relative polarizations.

To inspect the  current-induced dynamics of the  DW at $z=z_0$,
we evaluate the accumulated spin density that acts on
the DW at $z=z_0$. The DW magnetization dynamics are then
modeled using the Landau-Lifshitz equation
\footnote{The inclusion of Gilbert damping is
straightforward and has only a minor effect for the case of weak damping.}
\begin{equation}
\partial _t{\bf M}
=-\frac{\gamma J}{\sigma_{cs}}
{\bf M}\times {\bf S}[{\bf M}].
\label{magn}
\end{equation}
\begin{figure}
\includegraphics[width=0.45\textwidth]{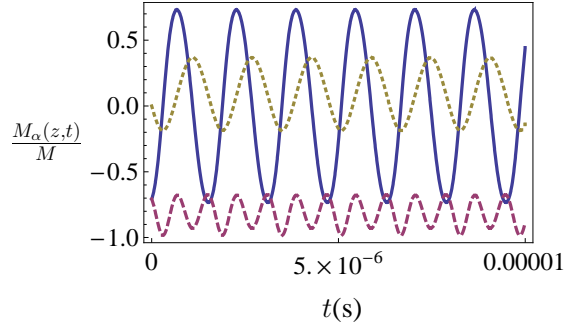}
\caption[$\vec{M}(t,\theta_0=\pi/4)$:Magnetization]
{(Color online) The time dependence
of the magnetization with the initial condition for the second wall to be
at an angle of $\theta_0=\pi/4$ to the first wall. This is the solution
to Eq.~\ref{magn}.  The solid
curve is the $x$-component, dashed  the $y$-component,
and dotted  the $z$-component. Taken at the centre of the domain wall.}\label{magthetapi4}
\end{figure}
As an initial condition we assume that the magnetization profile in the wire
without electric current is described by Eq.~(\ref{magnetization2}).
The results for the time dependence of the magnetization are shown in
Fig.~\ref{magthetapi4} for the centre
of the DW, $z=z_0$. We should note that the relative orientation of the
walls at the start of motion does play a role in the type of motion we see.
Here we present it for an arbitrary configuration. As
we move away from the centre of the DW, the relative orientation
becomes increasingly irrelevant. At $z=z_0+L$ the motion is the same
regardless of the value of $\theta_0$.

Analyzing  different magnetization components, we find
their motions have
different frequencies and a different form. No lateral movement or permanent distortion of the DW
is observed, we see only oscillations. At the edge of the DW wall there are small rotations of
the magnetization, regardless of the initial conditions. If we move far from the
DW then all the dynamics of the magnetization vanish.

This is in contrast to the case where we do not include the first domain wall. In this case there is no motion near the centre of the domain wall at all.  Furthermore the oscillations in the magnetization towards the edge of the domain wall are much slower (by several orders of magnitude) than exhibited here.

\begin{figure}
\includegraphics[width=0.45\textwidth]{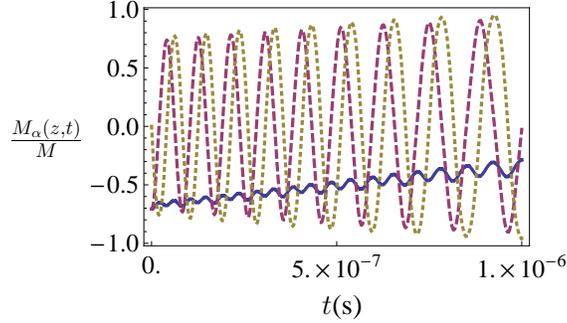}
\caption[$\vec{M}(t,\theta_0=\pi/4)$: Magnetization]{(Color online) Same as in Fig.~\ref{magthetapi4}
when the anisotropy is included, see Eq.~\ref{animagn}.  The solid
curve is the $x$-component, dashed  the $y$-component, and dotted
the $z$-component. Taken at the centre of the domain wall.}\label{anis}
\end{figure}

\begin{figure}
\includegraphics[width=0.45\textwidth]{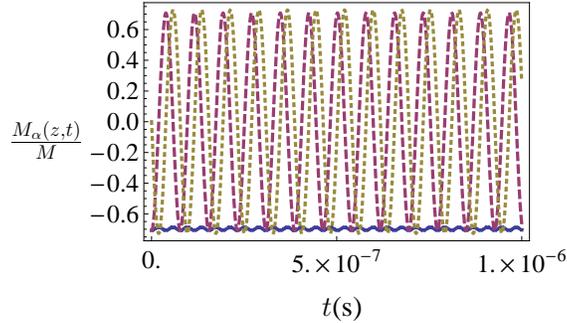}
\caption[$\vec{M}(t,\theta_0=\pi/4)$: Magnetization]{(Color online) The magnetization dynamics of the second DW without the influence of the first DW,
when the anisotropy is included, see Eq.~\ref{animagn}.  The solid
curve is the $x$-component, dashed  the $y$-component, and dotted
the $z$-component. Taken at the centre of the domain wall.}\label{aniszero}
\end{figure}

Extending our analysis to include the effect of magnetic anisotropy we write
\begin{eqnarray}
\label{animagn}
\partial_t{\bf M}
=-\frac{\gamma J}{\sigma_{cs}}{\bf M}\times {\bf S}[{\bf M}]
+\frac{\gamma K'}{M^2}{\bf M}\times\hat{{\bf x}}M_x.
\end{eqnarray}
We take the anisotropy constant $K'=-10$, and therefore the $x$-axis
as a hard magnetization axis.  Figure \ref{anis} shows the
effects of anisotropy on the domain wall motion. The anisotropy dampens motion in the x-direction, thus exacerbating the y and z oscillations.  This is also in contrast to the case where we ignore the first domain wall. In this case, although the anisotropy does introduce motion around the centre of the domain wall it does not involve a decaying $x$-component, see Fig.~\ref{aniszero}.

\section{Summary}

A current through a magnetic  nanowire containing DWs results
in a DW interaction mediated by the scattered charge carriers. We developed a method
for calculating the interaction energy and the consequences of this new coupling
mechanism. The DWs interaction energy oscillates as a function of the DWs mutual orientation
and distance. This has immediate consequences on how DWs rearrange upon applying a bias voltage
and on the fundamental limit of the DWs packing density. In fact, different
parts of the DW oscillate at different rates and in different ways:
becoming more regular, smaller, and quicker away from the DW centre.
The nonequilibrium DWs oscillations around the energy minima generates radiation with a frequency dependent on the applied bias voltage,
DW length and scattering strength. These
parameters are externally tunable for  utilizing  the interacting DWs as a versatile  radiation
source.
For an experimental realization  magnetic semiconductors \cite{ohno98,sugawara08,ruster03,koike05,fukumara05,holleitner05} are favorable, our results
 are in the range already achievable \cite{ruster03}.
 Extension to the metallic case is straightforward, though DW lengths may not be easily fabricated on the required scale. In this case the results remain qualitatively similar. Furthermore anisotropy will completely dampen any DW oscillations as motion in a plane becomes much harder due to the much larger magnetization size, $M=1.72\times10^{6}$Am${}^{-1}$ for Fe.

In our numerical simulations we used parameters of a magnetic semiconductor with relatively
large electron wavelength, $\lambda \leq L$, and much longer mean free path $l\gg \lambda $.
The latter can be realized in case of small density of impurities and defects.
However, magnetic semiconductors like GaMnAs are usually strongly disordered, and instead of a strong inequality one may find $l \geq \lambda $. In this case, the phase of the current-induced
spin density wave at a distance $z_0>l$ will be affected by impurities, and, therefore,
one can expect that the {\it disorder-averaged} interaction between two DWs at a distance $z_0$ is
suppressed by the factor $e^{-z_0/l}$. However, we should stress that the {\it real interaction}
between two DWs depends only on a given realization of the disorder and therefore
is {\it not damped} by the impurities. This effect is analogous to the nondamping
of the RKKY interaction between magnetic impurities in disordered metals\cite{zyuzin86,bulaevskii86}.
The detailed analysis of this phenomenon is beyond the scope of this paper.

\section*{Acknowledgments}
This work is supported by  DFG under  SPP 1165, the
FCT under PTDC/FIS/70843/2006 in Portugal, and by the Polish MNiSW
as a research project in years 2007 -- 2010.

\bibliography{references}

\end{document}